\documentclass[11pt]{article}
\usepackage[english]{babel}
\usepackage{graphicx}
\usepackage[authoryear,round]{natbib}
\usepackage{rotating}
\usepackage[margin=20mm]{geometry}
\usepackage{hyperref}
\hypersetup{
    colorlinks = true,
    linkcolor = blue,
    anchorcolor = blue,
    citecolor = blue,
    filecolor = blue,
    urlcolor = blue,
	pdfauthor={some author},
	pdftitle={eye-catching title}
    }
\usepackage{amssymb,amsmath}
\usepackage{url}

\begin{document}

\title{\emph{NuSTAR} view of the NGC~4151 corona:\\ thermal balance and astroparticle constraints}
\author{Luigi Foschini\\
\small{Brera Astronomical Observatory}\\
\small{National Institute of Astrophysics (INAF)}\\
\small{Milano/Merate, Italy}\\
\small{email: \texttt{luigi.foschini@inaf.it}}}
\date{November 16, 2026}
\maketitle

\begin{abstract}
I present a comprehensive analysis of all publicly available \emph{NuSTAR} observations of the nearby Seyfert galaxy NGC~4151 (2012–2024). The hard X-ray continuum is well described by thermal Comptonization in a hot corona, with electron temperatures $kT_{\rm e} \approx 40-125$~keV and optical depths $\tau \approx 0.4-1.2$ (slab geometry). A clear anti-correlation between $kT_{\rm e}$ and $\tau$ is observed, consistent with a thermostat behavior with constant Compton parameter $\gamma \approx 3.12$ calculated according \cite{ST80}. Variability on timescales of hours constrains the size of the emitting region to $R \approx 250-1200\,r_{\rm g}$, corresponding to moderate compactness values $\ell \lesssim 4$. The source occupies the same region of the compactness–temperature plane as other local Seyferts, with no evidence for pair runaway. I then explore two astroparticle implications. The \emph{IceCube} neutrino excess reported toward NGC~4151 corresponds to $L_\nu/L_X \approx 0.3\%$, an energetically negligible contribution that does not alter the observed coronal energy balance. An additional injection of pairs from the hypothetical X17 boson cannot produce significant changes, unless the injection rate exceeds the photoproduction one, but no anomalies were found in the present data set. The \emph{NuSTAR} data thus provide no evidence for either a significant neutrino cooling channel or an X17-related pair injection in the corona of NGC~4151.
\end{abstract}

\section{Introduction}
\label{intro}
The \emph{NuSTAR} satellite \citep{HARRISON2013}, launched in 2012, is the first instrument focusing X-rays above 10~keV. It is placed in low-Earth orbit, where the background is low. These two factors determined an improvement of the hard X-rays sensitivity up to two orders of magnitude with respect to previous hard X-rays satellites. On the focal plane, there are two CdZnTe pixellated detectors, named FPMA and FPMB (Focal Plane Module A, B). The nominal energy range is $3-78.4$~keV, with $0.4$~keV resolution at $10$~keV. Its superb performance makes it the best instrument available to date to study the hot corona of active galactic nuclei (AGN, e.g. \citealt{FABIAN2015}). 

NGC~4151 ($z=0.003326$, \citealt{PEDLAR1992}; $d_{\rm L}=15.8\pm0.4$~Mpc, \citealt{YUAN2020}) is one of the nearest, brightest, and best-studied Seyfert galaxies in hard X-rays (e.g. \citealt{ULRICH2000,LUBINSKI2010}), likely detected also at gamma rays (\citealt{DICOCCO1977}; see also \citealt{FOSCHINI2026} for a recent review on the gamma-ray detection). There is also an intriguing claim of a possible neutrino association ($\approx 3\sigma$, \citealt{ABBASI2025}). 

Over the past decades, several observations have constrained the parameters of its hot corona through thermal Comptonization models, clearly showing a thermostat behavior, with Compton parameter $y\approx 1$. Early hard X-ray observations searched for a non-thermal origin for the continuum. \cite{PACIESAS1977} reported a power-law photon index ($\Gamma \approx 1.6$), while a thermal-bremsstrahlung with temperature of $34$~keV resulted in a worst fit.  \cite{BAITY1984}, using \emph{HEAO-1} data up to $2$~MeV confirmed the previous results, although one epoch (June 1978) preferred a thermal model with $kT_{\rm e}\approx 85$~keV, where $k\approx 1.381\times 10^{-16}$~erg~K$^{-1}$ is the Boltzmann constant, and $T_{\rm e}$ is the electron temperature [K]. It is worth noting that the detection of gamma rays at MeV energies occurred during those years: \cite{DICOCCO1977} observation dates back to May 22-23, 1977; \cite{BAITY1984} detected it on December 1977 and December 1978, but not in June 1978, noting a break in the spectrum around 50~keV; \cite{PEROTTI1981} detected NGC~4151 again on September 30, 1979, while no detection was reported in May 17, 1980 \citep{PEROTTI1983}. It is therefore reasonable to assume that, during those years, there may have been an interplay between a thermal and a non-thermal component. It is worth citing the Fig.~2 in \cite{SZELECKA2019}, which shows that NGC~4151 had the highest X-ray flux ever detected just during those years.

\cite{TITARCHUK1994} applied a relativistic Comptonization model to joint \emph{Ginga} and \emph{CGRO}/OSSE spectra and found $kT_{\rm e} = 44^{+5}_{-3}$~keV, with optical depth depending on the geometry: $\tau = 1.25^{+0.16}_{-0.18}$ (disk geometry) or $\tau = 3.20^{+0.30}_{-0.37}$ (spherical geometry). Subsequent broad-band analyses reinforced the picture of a moderately optically thick corona. \cite{ZDZIARSKI1996} fitted \emph{ROSAT}, \emph{Ginga}, \emph{CGRO}/OSSE, and \emph{ASCA} data with thermal Comptonization, obtaining $kT_{\rm e} \approx 60$~keV and $\tau \approx 1$ (spherical geometry) after reflection was included. \cite{ZDZIARSKI2002} analyzed a long \emph{ASCA} observation jointly fitted with \emph{CGRO}/OSSE and derived $kT_{\rm e} \approx 70$~keV with an intrinsic photon index $\Gamma \approx 1.9$, implying $\tau \approx 1.3$. 

\cite{BECKMANN2005} presented the first \emph{INTEGRAL} spectrum of NGC~4151 and fitted a Comptonization model yielding $kT_{\rm e} = 94^{+4}_{-10}$~keV and $\tau = 1.3^{+0.2}_{-0.1}$. The most extensive study of flux-state dependence is that of \cite{LUBINSKI2010}, later extended by \cite{SZELECKA2019}, who analyzed \emph{INTEGRAL} data spanning bright and dim hard X-ray states. In the bright state they found $kT_{\rm e} \approx 50-70$~keV and $\tau \approx 1.3-2.6$, depending on the geometry of the corona; in the dim state $kT_{\rm e} \approx 180-230$~keV and $\tau \approx 0.3-0.7$. In both cases the Compton $y-$parameter remained $\approx 1$, indicating a stable geometry and thermostatic behavior.

Taken together, these measurements show that the corona of NGC~4151 has historically occupied the region $kT_{\rm e} \approx 40-100$~keV and $\tau \approx 1-2$ in bright states, with a tendency toward higher temperatures and lower optical depths in dim states, consistent with the thermostat behavior expected from pair-regulated or energy-balance Comptonization. Uncertainties remain large for the oldest data and for the dimmest states, where the high-energy cutoff is poorly sampled. 

\emph{NuSTAR} observed NGC~4151 many times, but there are only a few works published on a few specific observations (e.g. \citealt{KECK2015,BEUCHERT2017,SZANECKI2021,GIANOLLI2023}). Aim of the present work is to analyze the whole available \emph{NuSTAR} data set, to understand the behavior of the corona with the best data set available to date. I conclude by exploring the possibility to use this type of study in the field of astroparticle physics (neutrinos, X17 boson).

\section{NuSTAR observations}
As of September 15, 2026, there are 20 \emph{NuSTAR} publicly available observations within $4'$ from the coordinates of NGC~4151. One is approved, but yet to be done. Two have issue flag set to 1. Three have too short exposures. Therefore, I selected and downloaded the remaining 14 observations. The list of the observations analyzed in the present work is shown in Table~\ref{tab:nustar}.

\begin{table}[t]
\caption{List of \emph{NuSTAR} observations analyzed in the present work.}
\begin{center}
\begin{tabular}{clc}
\hline
ObsID & Date [UTC] & Exposure [ks]\\
\hline
60001111002 & November 12, 2012, $06:06:03$ & 21.9\\
60001111005 & November 14, 2012, $08:16:05$ & 61.5\\
60502017002 & July 24, 2019, $08:06:09$ & 31.7\\
60502017004 & November 12, 2019, $09:21:10$ & 43.7\\
60502017006 & December 24, 2019, $11:21:07$ & 32.4\\
60502017008 & January 10, 2020, $07:36:12$ & 30.6\\
60502017010 & January 19, 2020, $12:11:05$ & 29.7\\
60502017012 & January 23, 2020, $06:26:12$ & 28.8\\
60901003002 & December 16, 2022, $03:06:12$ & 97.0\\
60902010002 & December 27, 2023, $10:36:12$ & 45.2\\
60902010004 & May 19, 2024, $09:26:12$ & 46.0\\
61001014002 & May 21, 2024, $13:06:06$ & 24.5\\
61001014004 & May 24, 2024, $18:01:09$ & 33.6\\	
60902010006 & June 15, 2024, $12:11:05$ & 48.0\\
\hline
\end{tabular}
\end{center}
\label{tab:nustar}
\end{table}

Data were reduced and analyzed by using \texttt{HEASoft v. 6.37.1} and \texttt{CALDB} updated on the same date, following the standard procedures \citep{PERRI2024}. The source counts were extracted from a circular region with $60''$ radius, while the background region was still a circle with radius $120''-150''$, depending on the availability of a source-free zone. The spectra were then rebinned to have at least 30 counts per bin, so to apply the $\chi^2$ test statistic. Having noted some residual divergence between the low-energy spectra of FPMA and FPMB, I considered $4-78$~keV as the valid energy range to use for spectral fitting. For each ObsID, I also extracted the lightcurve in the $10-78$~keV energy band by using the same extraction regions and time bin set to $3600$~s. I subtracted the background, after having renormalized for the different extraction regions. 

\begin{table}[t]
\caption{Summary of the model parameters derived from the spectral fit. Columns: (1) Observation Identifier (see Table~\ref{tab:nustar}); (2) intrinsic absorption column [$10^{22}$~cm$^{-2}$]; (3) covering fraction; (4) Photon index; (5) electron plasma temperature [keV]; (6) optical depth (slab geometry, calculated from \citealt{ST80}); (7) centroid energy of the iron emission line [keV]; (8) $\chi^2$ and degrees of freedom (dof); (9) unabsorbed flux in the $10-100$~keV energy band [$10^{-10}$~erg~cm$^{-2}$~s$^{-1}$]. Errors are at 90\% confidence level ($1.645\sigma$).}
\centering
\vskip 12pt
\begin{tabular}{ccccccccc}
\hline
ObsID & $N_{\rm H}^{z}$ & CF & $\Gamma$ & $kT_{\rm e}$ & $\tau$ & $E_{\rm Fe}$ & $\chi^2$/dof & $F$\\{}
(1)   & (2) & (3) & (4) & (5) & (6) & (7) & (8) & (9)\\
\hline
60001111002 & $34\pm 6$ & $0.55\pm 0.03$ & $1.84\pm 0.04$ & $42_{-11}^{+24}$ & $1.1\pm 0.4$ & $6.30_{-0.04}^{+0.03}$ & $1404.25/1361$ & 7.36\\
60001111005 & $32\pm 3$ & $0.60\pm 0.02$ & $1.82\pm 0.02$ & $64_{-16}^{+31}$ & $0.81_{-0.15}^{+0.19}$ & $6.30_{-0.01}^{+0.04}$ & $2108.79/1964$ & 8.60\\
60502017002 & $30\pm 4$ & $0.66\pm 0.02$ & $1.81\pm 0.03$ & $73_{-13}^{+106}$ & $0.72_{-0.29}^{+0.09}$ & $6.26_{-0.02}^{+0.01}$ & $1615.24/1530$ & 7.37\\
60502017004 & $26\pm 4$ & $0.54\pm 0.03$ & $1.83_{-0.02}^{+0.03}$ & $89_{-57}^{+102}$ & $0.60_{-0.19}^{+0.36}$ & $6.29_{-0.01}^{+0.02}$ & $1755.91/1766$ & 8.84\\
60502017006 & $25\pm 5$ & $0.50_{-0.03}^{+0.04}$ & $1.86\pm 0.03$ & $79_{-26}^{+89}$ & $0.64_{-0.22}^{+0.16}$ & $6.28_{-0.01}^{+0.02}$ & $1627.05/1553$ & 8.08\\
60502017008 & $31\pm 5$ & $0.64\pm 0.03$ & $1.83\pm 0.04$ & $63_{-21}^{+63}$ & $0.78_{-0.27}^{+0.19}$ & $6.27_{-0.02}^{+0.01}$ & $1550.30/1449$ & 7.04\\ 
60502017010 & $30_{-4}^{+5}$ & $0.60_{-0.02}^{+0.03}$ & $1.81\pm 0.04$ & $79_{-31}^{+117}$ & $0.67_{-0.27}^{+0.19}$ & $6.26\pm 0.01$ & $1588.74/1453$ & 7.41\\
60502017012 & $27_{-6}^{+5}$ & $0.60_{-0.03}^{+0.04}$ & $1.80\pm 0.04$ & $125_{-70}^{+77}$ & $0.45_{-0.11}^{+0.23}$ & $6.30_{-0.01}^{+0.03}$ & $1489.76/1434$ & 7.48\\
60901003002 & $21_{-3}^{+2}$ & $0.70_{-0.03}^{+0.04}$ & $1.82\pm 0.02$ & $47_{-4}^{+2}$ & $1.03_{-0.03}^{+0.05}$ & $6.26_{-0.01}^{+0.02}$ & $2515.51/2216$ & 8.84\\
60902010002 & $25\pm 4$ & $0.55_{-0.02}^{+0.03}$ & $1.85\pm 0.02$ & $62_{-15}^{+27}$ & $0.79_{-0.14}^{+0.13}$ & $6.26_{-0.01}^{+0.06}$ & $2057.00/1848$ & 10.9\\
60902010004 & $26\pm 3$ & $0.69_{-0.02}^{+0.03}$ & $1.85\pm 0.03$ & $50_{-10}^{+16}$ & $0.93_{-0.13}^{+0.14}$ & $6.27_{-0.01}^{+0.04}$ & $1889.07/1769$ & 8.56\\ 
61001014002 & $34\pm 4$ & $0.75\pm 0.02$ & $1.90_{-0.05}^{+0.04}$ & $107_{-41}^{+143}$ & $0.39_{-0.38}^{+0.34}$ & $6.32\pm 0.02$ & $1452.64/1377$ & 6.90\\
61001014004 & $27\pm 4$ & $0.72_{-0.03}^{+0.04}$ & $1.80\pm 0.03$ & $40_{-8}^{+12}$ & $1.19_{-0.26}^{+0.27}$ & $6.29\pm 0.01$ & $1632.67/1602$ & 8.03\\
60902010006 & $31\pm 3$ & $0.73\pm 0.02$ & $1.86\pm 0.03$ & $43_{-8}^{+11}$ & $1.05_{-0.22}^{+0.23}$ & $6.27_{-0.01}^{+0.03}$ & $1741.38/1751$ & 7.97\\
\hline
\end{tabular}
\label{tab:fits}
\end{table}%

\subsection{Spectra}
\label{spec}
The X-ray spectrum of NGC~4151 is generally modeled with a Galactic absorption, an additional partial covering, a warm absorber, a thermal Comptonization plus reflection, and one or more emission lines of the iron complex, which can sometimes be included in the reflection models (see the references in Sect.~\ref{intro}). I used the ObsID $60001111002$, the first of the list, to test the possible models, searching for a baseline model to be applied to all the observations, so to have a homogeneous parameter set. Given the current data set, I found the best solution is composed of:

\begin{itemize}
\item the T\"ubingen-Boulder interstellar matter absorption model (\texttt{tbabs} in \texttt{xspec}, \citealt{WILMS2000}), fixed to the Galactic value ($N_{\rm H}^{\rm Gal}=2.07\times 10^{20}$~cm$^{-2}$, \citealt{HI4PI});

\item a partial covering fraction absorption (\texttt{zpcfabs} in \texttt{xspec}), located at NGC~4151 ($z=0.003326$, \citealt{PEDLAR1992}); 

\item no need to model the ionized absorber, because the low-energy threshold of \emph{NuSTAR} starts from 4~keV and the most common edges (O, Ne, Fe L,...) are in the soft X-rays; in any case, I have carried out some security checks to confirm this assumption;

\item thermal Comptonization model (\texttt{nthcomp} in \texttt{xspec}, \citealt{ZDZIARSKI1996,ZYCKI1999}); the \texttt{inp\_type} parameter was fixed to 1 (disk blackbody); the seed photons from the disk were generally assumed to be between 5~eV \citep{ZDZIARSKI2002} and 10~eV \citep{LUBINSKI2010}: I considered a reference value of 7~eV (fixed); in any case, varying the temperature of the disc within this range has a negligible effect on the parameters of the corona;

\item reflected power-law from a neutral medium (\texttt{pexrav} in \texttt{xspec}, \citealt{MAGDZIARZ1995}); the photon index $\Gamma$ is linked to that from \texttt{nthcomp}; the reflection scaling factor \texttt{rel\_refl} was set to $-1$ (fixed), corresponding to reflection only; the cutoff energy was set $E_{\rm c}=2.7kT_{\rm e}$, where $kT_{\rm e}$ is given by \texttt{nthcomp}; the inclination angle is set to $58^{\circ}$ \citep{BENTZ2022}, corresponding to $\cos i =0.53$ (fixed);

\item an unresolved Gaussian emission line (\texttt{zgauss} in \texttt{xspec}); \cite{ZDZIARSKI2002} reported $\sigma_{\rm Fe}\approx 0.05-0.1$~keV, two components, one originating on kpc scales, and the other closer to the black hole; given the spectral resolution of \emph{NuSTAR}, I set $\sigma_{\rm Fe}=0$ (unresolved, fixed).
\end{itemize}

The model \texttt{nthcomp} does not measure the optical depth $\tau$, but it is possible to derive it from $\Gamma$ and $kT_{\rm e}$ by following \cite{ST80}. The photon index is linked to the spectral index $\alpha = \Gamma - 1$ and, in turn, to the parameter $\gamma$ in the Eq.~(17) in \cite{ST80}:

\begin{equation}
\alpha = \sqrt{\gamma + \frac{9}{4}} - \frac{3}{2}
\label{eq:alpha}
\end{equation}

The Compton parameter $\gamma$ is defined as in the Eq.~(14) of \cite{ST80}:

\begin{equation}
\gamma = \frac{\pi^2 m_{\rm e}c^2}{12(\tau + \frac{2}{3})^2 kT_{\rm e}}
\label{eq:gamma}
\end{equation}

\noindent where $m_{\rm e}\approx 9.109\times 10^{-28}$~g is the electron rest mass, and $c\approx 2.998 \times 10^{10}$~cm~s$^{-1}$ is the speed of light in vacuum. It is worth noting that Eq.~(14) of \cite{ST80} has a factor of 3 in the denominator, because it refers to a spherical geometry, but I selected a slab as suggested by polarization observations \citep{GIANOLLI2023}, and therefore the factor becomes 12 (cf Sect.~II.4 in \citealt{ST80}). This parameter $\gamma$ defines the emergent X-ray spectrum and it is a more accurate version of the Compton parameter $y$, because it takes into account the geometry and radiative transfer. The summary of the spectral fits is shown in Table~\ref{tab:fits}.

\begin{figure}[t]
\begin{center}
\includegraphics[width=0.5\textwidth]{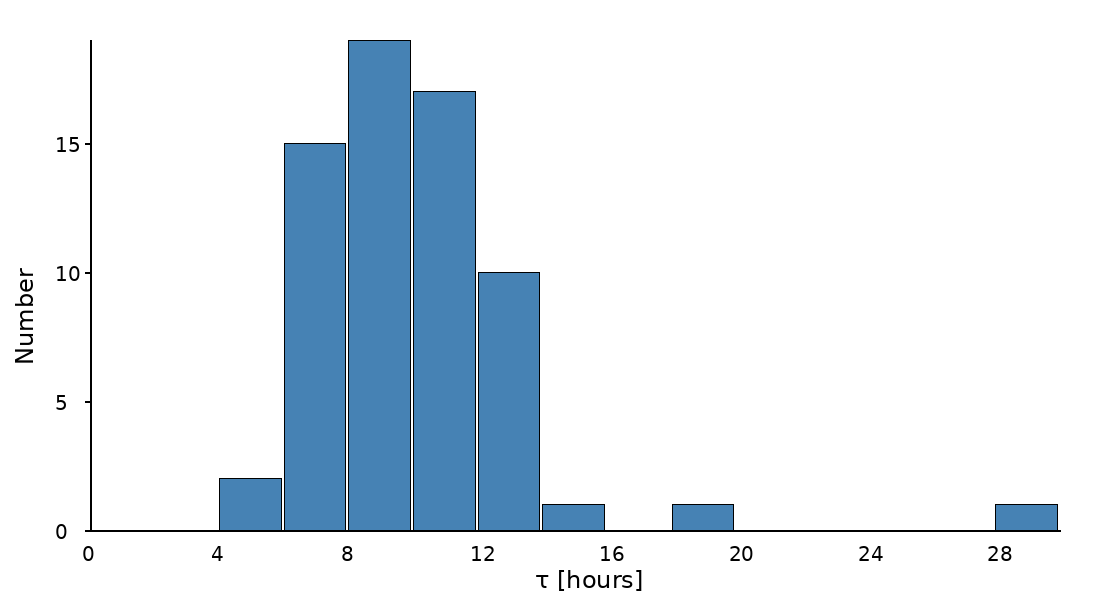}
\caption{Distribution of $\tau_{\rm var}$ measured from flux ratios greater than $3\sigma$.}
\label{fig:vardist}
\end{center}
\end{figure}

\begin{table}[!ht]
\caption{Summary of the variability analysis. Columns explanation: (1) Observation Identifier (the asterisk indicates measurements of $\tau_{\rm var}$ on larger bins); (2) Minimum doubling/halving time $|\tau_{\rm var}|$ [hours]; (3) X-ray flux extrapolated to the $0.1-200$~keV energy band [$10^{-9}$~erg~cm$^{-2}$~s$^{-1}$]; (4) Size of the emitting region [$r_{\rm g}$]; (5) Compactness parameter $\ell$.}
\centering
\vskip 12pt
\begin{tabular}{lcccc}
\hline
ObsID & $|\tau_{\rm var}|$ & $F$ & $R$ & $\ell$ \\
(1)   & (2)        & (3) & (4) & (5) \\         
\hline
60001111002  & $5.84\pm0.87$ & 2.10 & $256\pm 38$ & $2.71\pm 0.49$ \\
60001111005  & $6.25\pm0.92$ & 2.45 & $274\pm 41$ & $2.95\pm 0.53$ \\
60502017002  & $5.38\pm0.79$ & 1.93 & $236\pm 35$ & $2.70\pm 0.49$ \\
60502017004  & $7.63\pm1.20$ & 2.39 & $335\pm 53$ & $2.36\pm 0.45$ \\
60502017006  & $9.10\pm2.02$ & 2.24 & $399\pm 89$ & $1.85\pm 0.46$ \\
60502017008$^*$  & $28.0\pm5.3$  & 1.91 & $1230\pm 234$ & $0.51\pm0.11$ \\
60502017010$^*$  & $19.8\pm5.8$  & 1.74 & $868\pm 255$ & $0.66\pm0.21$ \\
60502017012  & $6.92\pm1.43$ & 1.96 & $304\pm 63$ & $2.13\pm 0.50$ \\
60901003002  & $6.70\pm0.96$ & 2.58 & $294\pm 42$ & $2.90\pm 0.52$ \\
60902010002 & $6.53\pm0.98$ & 3.20 & $286\pm 43$ & $3.68\pm 0.67$ \\
60902010004 & $6.73\pm0.98$ & 2.33 & $295\pm 43$ & $2.61\pm 0.47$ \\
61001014002$^*$ & $13.6\pm3.2$  & 2.07 & $596\pm 140$ & $1.15\pm 0.29$ \\
61001014004 & $9.12\pm1.80$ & 2.16 & $400\pm 79$ & $1.78\pm 0.40$ \\
60902010006 & $8.15\pm1.92$ & 2.28 & $358\pm 84$ & $2.10\pm 0.54$ \\
\hline
\end{tabular}
\label{tab:variability}
\end{table}

\subsection{Lightcurves}
I searched for doubling/halving characteristic time $\tau_{\rm var}$ between two consecutive points of the lightcurves:

\begin{equation}
\tau_{\rm var}=-\frac{t_1 - t_0}{\log_2 \frac{F(t_1)}{F(t_0)}}
\label{eq:timevar}
\end{equation}

\noindent where $F(t_0)$ and $F(t_1)$ are the fluxes (count rates) at the times $t_0$ and $t_1$, respectively, with $t_0<t_1$. The threshold on the flux difference is set to be greater than $3\sigma$. If no significant flux changes were found between two consecutive points, then I proceeded to consider flux changes between greater time bins. I found 66 significant flux changes, and the distribution of $\tau_{\rm var}$ is shown in Fig.~\ref{fig:vardist}. The values span from a minimum of $\approx 5.4$~h to a maximum of $\approx 28$~h, and a median value of $\approx 9.75$~h. Since in the present work it is not important to know if the flux is increasing ($\tau_{\rm var}$ negative) or decreasing ($\tau_{\rm var}$ positive), in the following I simply use $\tau_{\rm var}$ without the sign. 

From the characteristic time, it is possible to estimate the size of the emitting region:

\begin{equation}
R < \frac{c\tau_{\rm var}}{1+z}
\label{eq:size}
\end{equation}

These measurements are necessary to calculate the compactness parameter $\ell$ \citep{GUILBERT1983}:

\begin{equation}
\ell = \frac{\sigma_{\rm T}L_X}{R m_{\rm e}c^3}
\label{eq:compact}
\end{equation}

\noindent where $\sigma_{\rm T}\approx 6.652\times 10^{-25}$~cm$^{2}$ is the Thomson cross section. To estimate the corona luminosity $L_X$, I converted the count rates into physical units by using the spectral model of the same ObsID (cf Table~\ref{tab:fits}). Then, I extrapolated the flux of the power-law component in the $0.1-200$~keV energy band, as suggested by \cite{FABIAN2015}, and calculated the luminosity by using the distance $15.8$~Mpc \citep{YUAN2020}. To estimate the size of the emitting region $R$, I selected the minimum $\tau_{\rm var}$ of each ObsID. The resulting value is then converted in units of gravitational radius $r_{\rm g}\approx 2.5\times 10^{12}$~cm, calculated by using the mass of the central black hole $M=1.66_{-0.34}^{+0.48}M_{\odot}$ \citep{BENTZ2022}. The results are shown in Table~\ref{tab:variability}.

\begin{figure}[!t]
\begin{center}
\includegraphics[width=\textwidth]{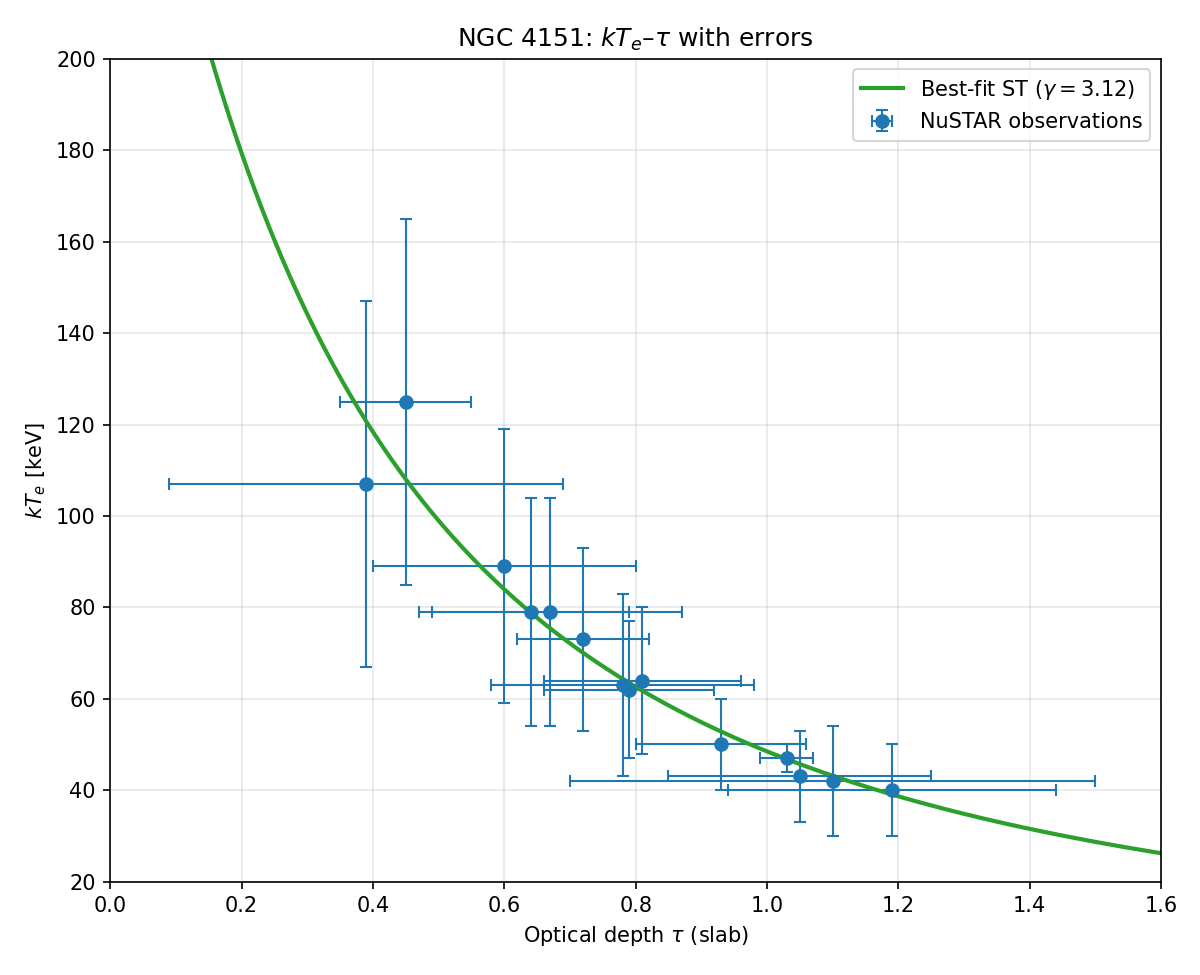}
\caption{Electron plasma temperature [keV] vs optical depth. Data points from the present \emph{NuSTAR} observations. The green continuous curve represents the best fit.}
\label{fig:summary1}
\end{center}
\end{figure}

\begin{figure}[!h]
\begin{center}
\includegraphics[width=\textwidth]{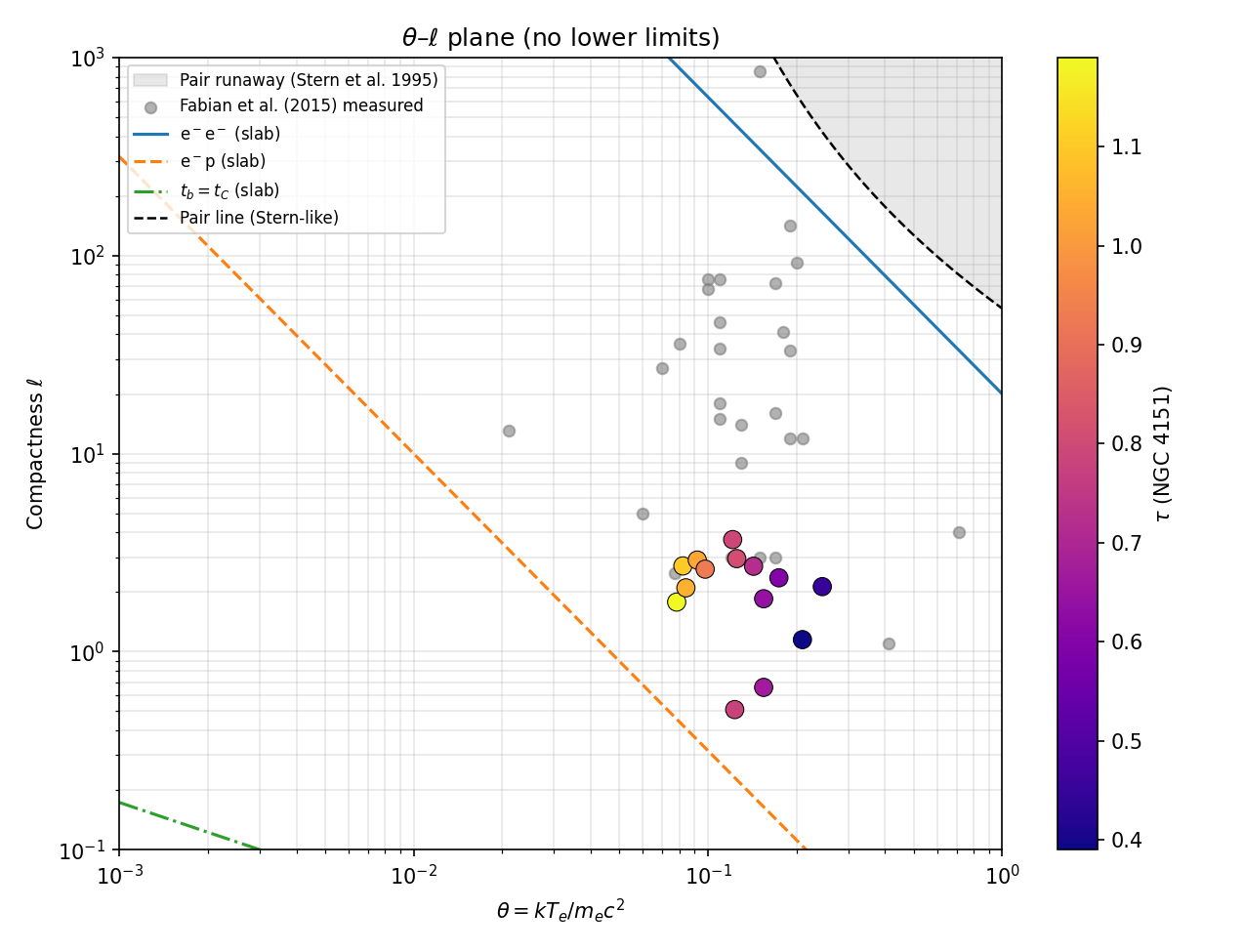}
\caption{Compactness $\ell$ vs electron plasma temperature $\theta$ in units of $m_{\rm e}c^2$. Colored data points are from the present set of \emph{NuSTAR} observations, and each color is associated to the optical depth $\tau$ according to the color bar on the right. The gray points are the data from \cite{FABIAN2015}, after having excluded the upper limits. The gray region is the pair runaway zone according to \cite{STERN1995}. The continuous, dashed, and dot-dashed lines are the boundaries for the electron-electron and electron-proton couplings, and the dominance of the Compton cooling, respectively, calculated according to \cite{FABIAN2015}. }
\label{fig:summary2}
\end{center}
\end{figure}

\section{Discussion}
\subsection{Overview}
The hard X-ray continuum is modeled with a thermal Comptonization component (hot electron plasma) that up-scatters soft disk photons, accompanied by a partially covering absorber and an unresolved iron emission line near $\approx 6.26–6.32$~keV (Table~\ref{tab:fits}). The photon index is stable at $\Gamma \approx 1.80–1.90$, typical of Seyfert 1 galaxies \citep{BECKMANN2005,RICCI2011,FABIAN2015}. The electron temperature $kT_{\rm e}$ ranges from $\approx 40$~keV to $\approx 125$~keV, while the optical depth $\tau$ (computed in the \citealt{ST80} formalism for slab geometry, preferred according to \citealt{GIANOLLI2023}) varies between $\approx 0.39$ and $\approx 1.19$. The unabsorbed $10–100$~keV flux is $F \approx (7-11)\times10^{-10}$~erg~cm$^{-2}$~s$^{-1}$, showing moderate variability by a factor of $\approx 1.5$. The intrinsic column density is $N_{\rm H}^{z} \approx (21-34)\times10^{22}$~cm$^{-2}$, with covering fractions CF $\approx 0.5–0.75$, consistent with the well-known partially covering absorber in this source. Fit quality is acceptable ($\chi^{2}/$dof $\approx$ 1.0–1.1).

A clear anti-correlation is observed between electron temperature and optical depth: hotter coronae are optically thinner, and vice versa (Figure~\ref{fig:summary1}). The best-fit relation ($\gamma \approx 3.12$) is consistent with the expected heating/cooling balance in Comptonizing plasmas. This behavior is standard for the coronae of Seyfert AGN and matches the trends reported in earlier observations and larger samples (see the references in Sect.~\ref{intro}).

Minimum doubling/halving times are $\tau_{\rm var} \approx 5–28$~hours (Table~\ref{tab:variability}). Using the light-crossing argument and the $0.1–200$~keV luminosity, the size of the emitting region is $R \approx 250–1200 r_{\rm g}$. The corresponding compactness parameter spans $\ell \approx 0.5–3.7$. These sizes are characteristic of a compact corona (smaller than hundreds of gravitational radii) rather than a structure extended on the scale of the outer disk.

Fig.~\ref{fig:summary2} shows the NGC~4151 measurements (colored by optical depth $\tau$) lie in the same region of the compactness–temperature plane occupied by other AGN from the \cite{FABIAN2015} sample, well below the pair-runaway line of \cite{STERN1995}. There is no evidence for electron–positron pair runaway. The data are consistent with an electron-dominated plasma in which Compton cooling is efficient, lying between the theoretical $e^{-}$–$e^{-}$ and $e^{-}$–$p$ equilibrium curves. 

To conclude, the corona of NGC~4151 follows the standard picture for Seyfert galaxies: a hot thermal plasma ($kT_{\rm e} \approx 40–125$~keV) with moderate optical depth ($\tau \approx 0.4–1.2$), a heating/cooling balance that produces the observed $kT_{\rm e}$–$\tau$ anti-correlation, and moderate compactness ($\ell \lesssim 4$) insufficient to trigger pair runaway. The emitting region is moderately compact (hundreds of $r_{\rm g}$) and varies on hourly timescales, consistent with a central corona with confined extended geometry. No significant deviations from the broader AGN population are found (e.g.  \citealt{BECKMANN2009,RICCI2011,FABIAN2015}): NGC 4151 is a prototypical classical corona exhibiting the usual state variability (temperature versus optical depth) without extreme non-thermal signatures or pair saturation.

\subsection{Astroparticle physics: neutrino}
\label{neutrino}
I examine the energetic implications of the neutrino excess reported by \emph{IceCube} in the direction of NGC~4151 \citep{ABBASI2025} in the context of the coronal properties derived from the present full set of \emph{NuSTAR} observations. Indeed, the neutrino would be an additional cooling channel, which may affect the energy balance of the corona. Since X-ray observations do not show any deviation from the standard heating/cooling balance, the neutrino detection might be in tension.

\emph{IceCube} measured a best-fit muon-neutrino flux of $\phi_{\nu} = 1.51^{+0.99}_{-0.81}\times10^{-11}$~TeV$^{-1}$~cm$^{-2}$~s$^{-1}$ at 1~TeV with spectral index $\gamma_{\nu} = 2.83^{+0.35}_{-0.28}$ \citep{ABBASI2025}. The bulk of the contribution to the test statistic arises from the $4.3–65$~TeV band. Adopting a luminosity distance of 15.8~Mpc \citep{YUAN2020} and integrating the power-law spectrum over this energy range, I obtain an isotropic muon-neutrino luminosity $L_\nu \approx 2.3\times10^{41}\,\text{erg s}^{-1}$.

From the present \emph{NuSTAR} analysis, the unabsorbed $0.1–200$~keV flux is typically $F_X\approx(1.7-3.2)\times10^{-9}$~erg~cm$^{-2}$~s$^{-1}$. A representative value $F_X\approx 2.5\times10^{-9}$~erg~cm$^{-2}$~s$^{-1}$ corresponds to $L_X \approx 7.5\times10^{43}\,\text{erg s}^{-1}$, so that:

\begin{equation}
\frac{L_\nu}{L_X} \approx 3\times10^{-3}
\label{eq:neutrino}
\end{equation}

This estimate is the lowest among recent calculations for NGC 4151 and for the nearby Seyfert 2 NGC 1068 \citep{FIORILLO2024A,FIORILLO2024B,NERONOV2024,TESTAGROSSA2026}, but remains compatible with them within an order of magnitude. Differences of this order arise from the choice of X-ray flux (either from a specific energy band or from a spectral energy distribution), the treatment of absorption and reflection, and the specific \emph{IceCube} data reduction and neutrino flavor estimates. The model of \cite{INOUE2023} predicted a neutrino flux about two orders of magnitude below the all-flavor flux corresponding to the \cite{ABBASI2025} measurement. When their coronal model is normalized to the observed \emph{IceCube} flux, the resulting $L_\nu/L_X$ ratio becomes $\approx0.3$, higher than the present estimate.

\emph{NuSTAR} observations show that the corona is maintained in thermal equilibrium by a balance between heating (magnetic dissipation or reconnection) and Compton cooling. An additional neutrino cooling channel at the level of $\lesssim 0.3\%$ of the X-ray luminosity is negligible with the current statistics. It cannot significantly alter the electron temperature, optical depth, or the Compton $\gamma$ parameter. This is consistent with the present results: the measured $kT_{\rm e}$--$\tau$ anti-correlation follows the standard constant $\gamma \approx 3.12$ locus, and the source occupies the same region of the compactness--temperature plane as the broader Seyfert sample \citep{FABIAN2015}, with no evidence of extra cooling. In the $\theta$--$\ell$ plane, NGC 4151 lies close to the $p$--$e^-$ coupling line, in the region where Compton cooling still dominates (see Fig.~\ref{fig:summary2}). This is consistent with the absence of a significant energetic contribution from hadronic cascades in the \emph{NuSTAR} band, in line with models in which the electromagnetic cascade associated with non-thermal protons remains subdominant relative to thermal Comptonization at hard X-ray energies \citep{TESTAGROSSA2026}.

It should be noted that the neutrinos primarily cool the non-thermal proton population rather than the thermal electrons responsible for the X-ray continuum. The two populations are only weakly coupled, so a neutrino luminosity at the percent level or below is not expected to leave a detectable imprint on $kT_{\rm e}$ or $\tau$. The present data therefore do not exclude the existence of a non-thermal proton population; they simply indicate that any associated energetic feedback in the hard X-ray band is currently below the sensitivity of \emph{NuSTAR}.

Interestingly, NGC~4151 was detected at MeV energies in the late 1970s \citep{DICOCCO1977,PEROTTI1981,PEROTTI1983,BAITY1984}. The corresponding luminosity was one to two orders of magnitude above the present neutrino luminosity. A MeV detection above the coronal cutoff is direct evidence of a non-thermal particle population (either primary non-thermal electrons or secondary pairs from hadronic cascades), because thermal Comptonization cannot produce significant emission at those energies. Contemporaneous \emph{COS-B} upper limits \citep{POLLOCK1981} in the 50~MeV--5~GeV band require a spectral break or steepening above a few MeV, ruling out a hard spectrum extending to GeV energies typical of a relativistic jet. This is consistent with a local non-thermal component (e.g., a hadronic cascade) that does not escape to GeV energies. Those observations may indicate a temporarily brighter hadronic cascade, in line with the expectation that the electromagnetic signature of non-thermal protons emerges primarily above the coronal cutoff. The historical hard X-ray light curve shows that the source was brighter in the past \citep{SZELECKA2019}, so a higher non-thermal proton activity at that epoch cannot be excluded. The MeV band therefore remains the most promising window to search for this component.

To summarize, even with the best hard X-ray instrument currently available (\emph{NuSTAR}) and one of the brightest Seyfert 1 galaxies in the hard X-ray band, no trace of an additional neutrino cooling channel is detected in the coronal energy balance. This further suggests that new X-ray observatories, which will allow for the measurement of coronal luminosity and spectral parameters with a precision of better than 1\%, could potentially constrain or detect energy contributions as small as the neutrino channel -- amounting to approximately 0.3\% -- discussed here. \href{https://www.cosmos.esa.int/web/athena}{\emph{NewAthena}} will offer a large effective area and high spectral resolution in the $0.2$--$12$~keV band, potentially improving constraints on the coronal continuum below the cutoff. However, the absence of hard X-ray coverage limits its ability to directly measure the high-energy cutoff of the corona. In any case, should its in-orbit performance be confirmed and systematics under control, studying the energy budget of the X-ray coronae of bright Seyfert galaxies could help identify new potential neutrino emitters.

\subsection{Astroparticle physics: X17 boson} 
For several decades now, efforts have been made to find experimental and/or observational evidence for dark matter, but without convincing results (see \citealt{CIRELLI2026} for an extensive and recent review). Among the different possibilities, the so-called X17 boson deserves some further attention. \cite{ATOMKI} reported an anomaly in the decay of the excited $^{8}$Be ($17.6-18.15$~MeV): the generated pairs show a significant ($\approx 7\sigma$) excess at $\Theta\approx 140^{\circ}$, being $\Theta$ the measured angular separation between electrons and positrons. Such result was interpreted as the $^{8}$Be decayed by emitting a light slow boson, which is soon converted into an electron-positron pair. Since the energy was $\approx 17$~MeV and the particle unknown, this candidate boson was named X17. If confirmed, it could open interesting possibilities to explain dark matter, being ``protophobic'', i.e. with negligible interaction with protons (e.g. \citealt{FENG2016,GUSTAVINO2024}). 

The early attempts to confirm the new particle with NA64 \citep{ANDREEV2021} and MEG~II \citep{AFANACIEV2025} experiments failed, while \cite{ANH2024} succeeded. However, the team of the latter experiment included also people of the Krasznahorkay's group, and therefore it cannot be considered an independent confirmation. The PADME experiment reported a promising weak excess ($\approx 2\sigma$) at $\approx 16.9$~MeV \citep{BOSSI2025}, so it is necessary to collect more data to understand if X17 is confirmed or not. 

As far as I am aware, very little has been done so far on an astrophysical scale, and what little has been done is essentially theoretical. \cite{WONG2020} proposed that self-gravitating assemblies of X17 (interpreted as isoscalar quantum electrodynamic mesons) could form compact objects that may produce a soft $\gamma$-ray (or $e^+e^-$) excess, unless they are sufficiently compact that gravitational binding suppresses the decays. \cite{KANAKIS2024} proposed that X17 may affect the equation of state of neutron or quark stars. \cite{ENBERG2025} set new constraints on X17 by using the \emph{IceCube} data on non-standard interactions.

Regardless of whether or not X17 might be a good candidate for solving the dark matter problem, I believe that—if confirmed—it could open up some very interesting physics. In this work, I therefore sought to determine whether any effects might be observable on astrophysical scales. 
Given that the estimated decay time of X17 is on the order of a picosecond \citep{ATOMKI,ANDREEV2021}, it is not even worth considering the construction of an instrument for direct observation. Since X17 decays into electron-positron pairs with an energy of around 8~MeV, one can attempt to assess what anomalous effects might arise from the additional injection of these pairs into the corona of NGC~4151. 

Pairs produced by the decay of the hypothetical X17 boson ($m_{\rm X17} c^2 \approx 17$ MeV) are injected into the corona with a kinetic energy of order $\approx 8.5$~MeV for each electron or positron. Their subsequent thermalization with the ambient coronal plasma ( $kT_e \approx 40-125$~keV) is governed by Coulomb collisions. The electron–electron (or electron–positron) energy exchange time can be estimated as \citep{SVENSSON1999}:

\begin{equation}
t_{\rm th} = 4 \frac{\theta^{3/2}(\sqrt{\pi}-1.2\theta^{1/4}+2\sqrt{\theta})}{n_{\rm e}\sigma_{\rm T}c\ln\Lambda}
\label{eq:coulomb}
\end{equation}

\noindent where $\ln\Lambda$ is the Coulomb logarithm. To estimate the pair density by photoproduction one can start from the energy density of a slab corona with the height $H$ much smaller than the radius $R$:

\begin{equation}
u_{\rm rad}\approx \frac{L_{X}}{\pi R^2 c (1+\tau)}
\label{eq:energydens}
\end{equation}

Under equilibrium condition, the pair density is linked to the photon density and to the corona temperature $\theta$, to take into account that only a fraction of photons with sufficient energy can produce pairs,  i.e. $n_{\pm}m_{\rm e}c^2\approx u_{\rm rad}\theta$. By combining the two equations, it results:

\begin{equation}
n_{\pm}\approx \frac{u_{\rm rad}\theta}{m_{\rm e}c^2} = \frac{L_{X}\theta}{\pi R^2 m_{\rm e}c^3 (1+\tau)}
\label{eq:paird}
\end{equation}

From the values of Tables~\ref{tab:fits} and \ref{tab:variability}, it results that $n_{\pm}\approx 5.6\times 10^6-2.3\times 10^8$~cm$^{-3}$, $\ln \Lambda \approx 29-31$, and $t_{\rm th} \approx 0.82-24\,\text{h}$. This is comparable or shorter than the light-crossing time of the corona $t_{\rm cross} = \tau_{\rm var}/(1+z) \approx 5.4-28$~h, and greater than the Compton cooling time $t_{\rm C} \approx 0.07-0.45$~h (calculated according to \citealt{FABIAN2015} and modified for a slab geometry with $H<<R$). 

\cite{SVENSSON1999} also considers the synchrotron self-absorption as a process to thermalize pairs, which requires that the magnetic compactness $\ell_{\rm B}$:

\begin{equation}
\ell_{\rm B}= \frac{\sigma_{\rm T}R u_{\rm B}}{m_{\rm e}c^2}\gtrsim 1
\end{equation}

\noindent where $u_{\rm B}=B^2/8\pi$ is the magnetic energy density [erg~cm$^{-3}$] and $B$ is the magnetic field [G]. Taking into account the size of the emitting region from Table~\ref{tab:variability}, the required magnetic field is $\approx 145-330$~G to be compared with the estimate of $\approx 30$~G by \cite{INOUE2023}. 

\begin{figure}[t]
\begin{center}
\includegraphics[width=\textwidth]{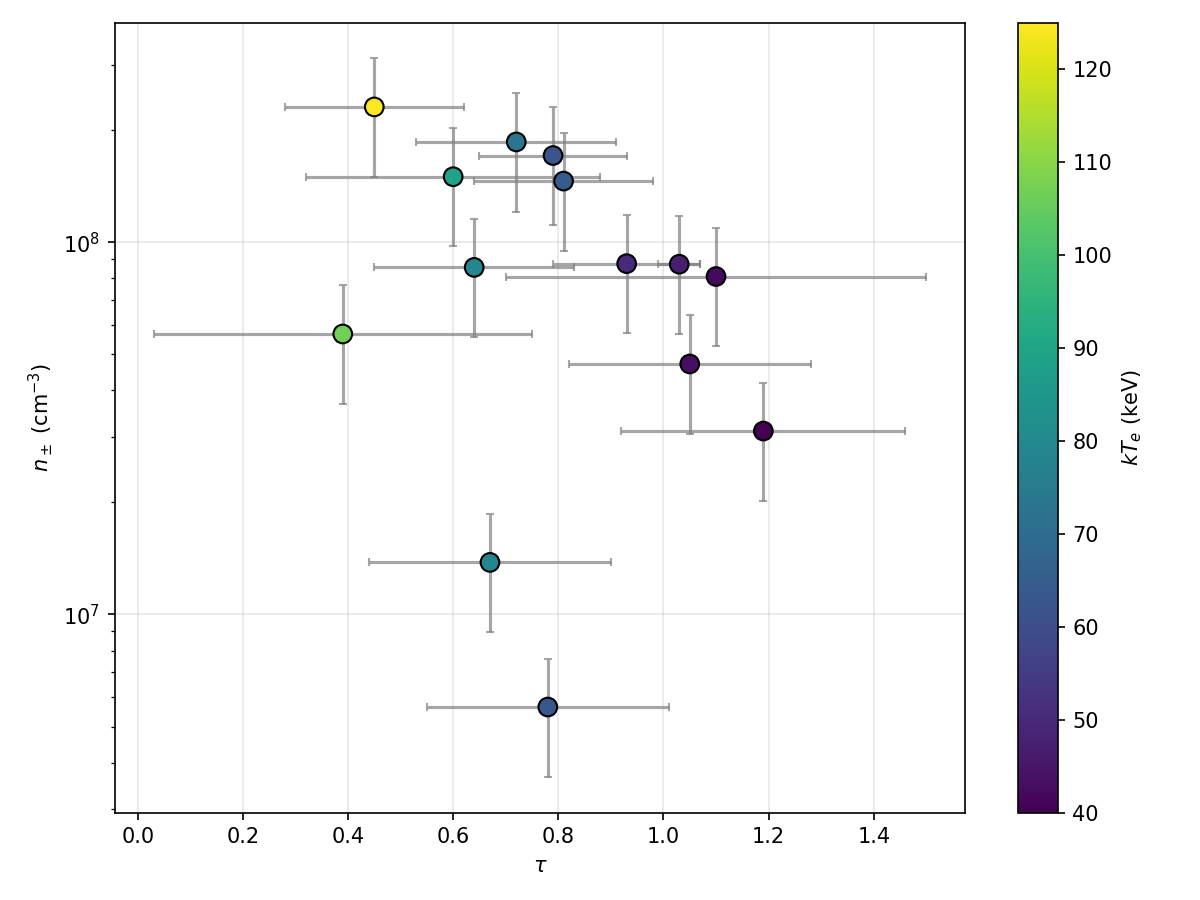}
\caption{Photoproduced pair density [cm$^{-3}$] as a function of the optical depth for each observation. Colors represent the electron temperature $kT_{\rm e}$ [keV].}
\label{fig:pairdens}
\end{center}
\end{figure}

Fig.~\ref{fig:pairdens} shows the photoproduction pair density (cf Eq.~\ref{eq:paird}) as a function of optical depth. The estimates are almost consistent each other within the errors, with only two outliers. Excluding the two outliers, the weighted mean is $n_{\pm}= (5.8\pm 0.7)\times 10^7$~cm$^{-3}$. The two outliers are ObsIDs 60502017008 and 60502017010, separated by 9 days (cf Table~\ref{tab:nustar}). The preceding observation (60502017006) was done 17 days earlier, while obsID 60502017012 was done after 4 days. So, I consider the latter for comparison. Within the uncertainties, the electron temperature and optical depth remain consistent across ObsIDs 60502017008, 60502017010 and 60502017012, while the size of the emitting region and the compactness change significantly. The corona contracts from a highly extended state in ObsID 60502017008 ($R \approx 1230\,r_g$, $\ell \approx 0.5$) to a compact configuration in ObsID 60502017012 ($R \approx 300\,r_g$, $\ell \approx 2.1$) over 13 days, with the intermediate pointing (ObsID 60502017010) occurring 9 days after the first and 4 days before the last. A weighted average of the two outliers yields $n_{\pm}=(6.8\pm 1.8)\times10^6$~cm$^{-3}$, compared with $n_{\pm}=(2.3\pm 0.8)\times10^8$~cm$^{-3}$ in the subsequent compact state, corresponding to an increase by a factor of $\approx 34$. This indicates that the variation is primarily geometric, with the microphysical properties of the plasma ($kT_e$ and $\tau$) staying essentially unchanged within the measurement errors. The lower pair density measured in the extended states is therefore attributable to the larger coronal size (and the correspondingly lower photon density), rather than to a change in the thermal state of the plasma.

To summarize: the Compton cooling time (4–27 minutes) is shorter than both the Coulomb thermalization time of X17 pairs (0.8–24 hours) and the observed variability timescale of the corona (5–28 hours), implying that any additional pair injection from X17 is rapidly compensated by the coronal energy balance. The magnetic compactness of the corona is too low for synchrotron self-absorption to contribute significantly to thermalization. One can try to search for some deviations from the relationship $kT_{\rm e}-\tau$ on shorter timescales. However, looking at the measurement errors and the exposures, it seems to be not possible. From Table~\ref{tab:nustar}, the shortest exposure is the first one, obsID 60001111002, with $21.9$~ks ($\approx 6.1$~h). The error on the measured $\tau$ is 36\%, while that on $kT_{\rm e}$ is about 57\%. With such errors, the extra pairs injected by X17 should be $n_{\pm}\gtrsim 3\times 10^9$~cm$^{-3}$ to affect the measurement of $\tau$ (at $3\sigma$), greater than the average photoproduction pair density. In addition, the estimated characteristic times are: $\tau_{\rm var}\approx 5.8$~h, $t_{\rm th}\approx 0.95$~h, and $t_{\rm C}\approx 0.084$~h. Therefore, a hypothetical temporary increase of pair density -- thus also $\tau$ -- is soon compensated by the energy balance of the corona. 

Therefore, it is necessary to search on shorter timescales. I have tried to extract the spectrum from the observation with the highest flux and short time, to have the best statistics. I selected the obsID 60902010002, which has the highest average flux $\approx 1.09\times 10^{-9}$~erg~cm$^{-2}$~s$^{-1}$. The characteristic times are: $\tau_{\rm var}\approx 6.5$~h, $t_{\rm th}\approx 0.82$~h, and $t_{\rm C}\approx 0.062$~h. Inspecting the lightcurve (not shown), it is possible to see that the peak bin is the latest one, recorded on December 28, 2023, between $12:15:55$ and $13:15:55$~UTC. I extracted the spectrum of that specific bin. Although the elapsed time is one hour, the effective exposure is $\approx 0.47$~h. I fit with the baseline model (see Sect.~\ref{spec}), and found these parameters: $\Gamma=1.83\pm 0.10$, $kT_{\rm e}<174$~keV, from which is calculated $\tau\gtrsim 0.2$. Therefore, even with the highest flux, there is not enough statistics to constrain $\tau$ on hourly timescales.

The contribution of extra pairs from X17 to the observed Thomson optical depth is therefore expected to be limited, unless the injection rate substantially exceeds the photoproduction rate. If the injection is continuous, there is no way to detect any clues; if, on the other hand, the injection occurs in pulses, then it becomes a matter of luck: one must manage to catch the corona in a state of temporary non-equilibrium. Anyway, the present \emph{NuSTAR} measurements of NGC 4151 thus exclude a pair-injection channel from X17 at a level capable of appreciably altering the observed coronal parameters.

\section{Final Remarks}
I have analyzed the full set of \emph{NuSTAR} observations of NGC~4151 available as of September 15, 2026. The corona behaves as a standard thermal Comptonizing plasma: electron temperature and optical depth are anti-correlated, the Compton $\gamma$-parameter remains of order unity, and the compactness is moderate. No evidence is found for pair runaway or for significant deviations from the properties of other local Seyfert coronae. 

Two astroparticle channels have been examined. The neutrino luminosity implied by the \emph{IceCube} excess is only $\approx 0.3\%$ of the coronal X-ray luminosity and is therefore energetically negligible; it cannot affect the observed heating/cooling balance. An extra pair injection from X17 is unlikely to affect the observed parameters, unless its rate exceeds photoproduction and it happens in bursts. The two outliers in pair density (cf Fig.~\ref{fig:pairdens}) are explained by a change in coronal size

These results demonstrate that high-quality hard X-ray spectroscopy and variability studies can place meaningful constraints on extra cooling or pair-injection channels in AGN coronae. Future instruments with larger effective area and higher spectral resolution, such as \emph{NewAthena}, will be able to push these limits below the $0.1\%$ level, potentially detecting or firmly excluding subtle energetic contributions from neutrinos or other non-standard processes.

\section*{Acknowledgements}
This research has made use of data and software provided by the High Energy Astrophysics Science Archive Research Center (HEASARC), which is a service of the Astrophysics Science Division at NASA/GSFC. I would like to thank Federico Testagrossa for his helpful comments on Sec.~\ref{neutrino}, which allowed me to provide clarification and highlight certain points more clearly.

\end{document}